\documentclass[twocolumn]{aastex631}

\usepackage[utf8]{inputenc}
\usepackage{amssymb}
\usepackage{newunicodechar}

\usepackage{amsmath}

\begin{document}

\title{Magnetic fields in the Horsehead Nebula}

\author[0000-0001-7866-2686]{Jihye Hwang}
\email{hjh3772@gmail.com}
\affil{Korea Astronomy and Space Science Institute (KASI), 776 Daedeokdae-ro, Yuseong-gu, Daejeon 34055, Republic of Korea}
\affil{University of Science and Technology, Korea (UST), 217 Gajeong-ro, Yuseong-gu, Daejeon 34113, Republic of Korea}
\affil{Department of Physics and Astronomy, University College London, Gower Street, London WC1E 6BT, UK}

\author[0000-0002-8557-3582]{Kate Pattle}
\affil{Department of Physics and Astronomy, University College London, Gower Street, London WC1E 6BT, UK}

\author{Harriet Parsons}
\affil{East Asian Observatory, 660 N. A'oh\={o}k\={u} Place, University Park, Hilo, HI 96720, USA}

\author{Mallory Go}
\affil{Brown University, Providence, RI 02912, USA}
\affil{Molokai High School, 2140 Farrington Ave, Ho'olehua, HI 96729, USA}

\author[0000-0002-1229-0426]{Jongsoo Kim}
\affil{Korea Astronomy and Space Science Institute (KASI), 776 Daedeokdae-ro, Yuseong-gu, Daejeon 34055, Republic of Korea}
\affil{University of Science and Technology, Korea (UST), 217 Gajeong-ro, Yuseong-gu, Daejeon 34113, Republic of Korea}

\begin{abstract}

 We present the first polarized dust emission measurements of the Horsehead Nebula, obtained using the POL-2 polarimeter on the Submillimetre Common-User Bolometer Array 2 (SCUBA-2) camera on the James Clerk Maxwell Telescope (JCMT). The Horsehead Nebula contains two sub-millimeter sources, a photodissociation region (PDR; SMM1) and a starless core (SMM2). We see well-ordered magnetic fields in both sources. We estimated  plane-of-sky magnetic field strengths of 56$\pm$9 and 129$\pm$21 $\mu$G in SMM1 and SMM2, respectively, and obtained mass-to-flux ratios and Alfv\'en Mach numbers of less than 0.6,  suggesting that the magnetic field can resist gravitational collapse and that magnetic pressure exceeds internal turbulent pressure in these sources. In SMM2, the kinetic and gravitational energies are comparable to one another, but less than the magnetic energy. We suggest a schematic view of the overall magnetic field structure in the Horsehead Nebula. Magnetic field lines in SMM1 appear have been compressed and reordered
during the formation of the PDR, while the likely more-embedded SMM2 may have inherited its field from that of the pre-shock molecular cloud. The magnetic fields appear to currently play an important role in supporting both sources.

\end{abstract}


\section{Introduction} \label{sec:introduction}

 Massive stars emit strong ultra-violet (UV) radiation, which ionizes their parental molecular clouds, creating HII regions (e.g., \citealt{Arthur2011, Walch2012}). This radiative feedback can also help to trigger further star formation by compressing the gas and dust surrounding massive stars (\citealt{Walch2012}).  The compressed neutral gas structures bounding HII regions are known as photodissociation regions (PDRs). PDRs are dense, often shell-like, cloud structures, with elongated dense structures protruding into the HII region, typically referred to as `pillars' or `elephant trunks'. These  pillars are created as the shock from the expanding HII region is driven into the surrounding molecular cloud, although their formation mechanism remains under debate (e.g., \citealt{Arthur2011}). Magnetic fields within PDRs may play an important role in supporting their structures against thermal pressure from the surrounding HII region or against collapse under self-gravity (e.g., \citealt{Pattle2018}). However, to date there have been only a few studies which have observed and analyzed magnetic fields within PDRs.

\begin{figure*}[htb!]
\epsscale{0.9}
\plotone{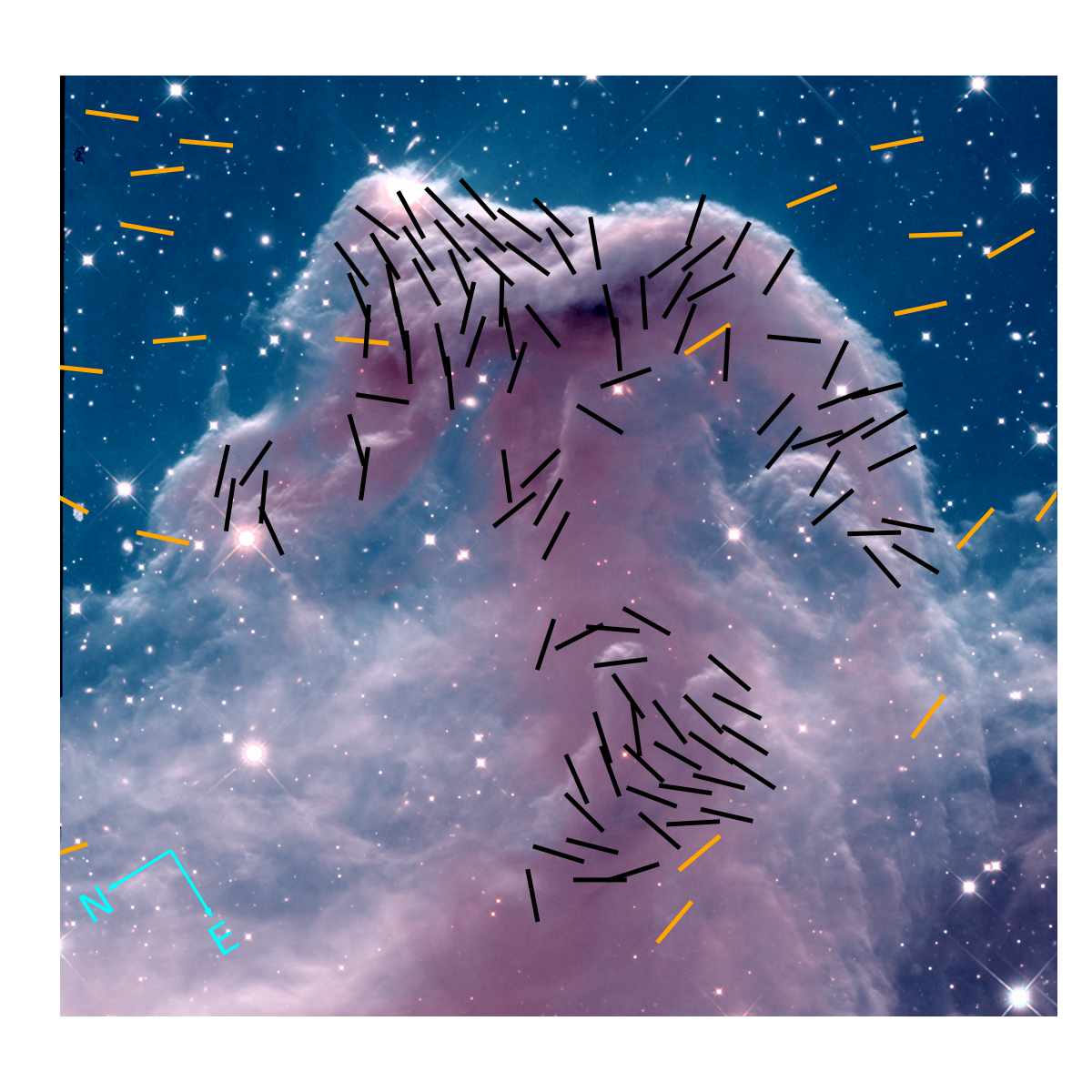}
\caption{Magnetic field vectors overlaid on a two-color composite of HST imaging in two broadband filters, F110W (YJ) and F160W (H), centered at 1153.4 and 1536.9 nm respectively, taken from the Mikulski Archive for Space Telescopes (MAST)\footnote{\url{https://archive.stsci.edu/prepds/heritage/horsehead/}}. Black and orange segments show magnetic field orientations inferred from JCMT POL-2 850 $\mu$m measurements (this work) and Palomar Observatory $r$-band measurements centered at $\sim$ 7200 \AA\ \citep{Zaritsky1987}, respectively. Note that the image is rotated with respect to standard equatorial coordinates.
\label{fig:hst}}
\end{figure*}

The Horsehead Nebula, shown in Figure \ref{fig:hst}, is one of the most well-known PDRs. Also known as B33, \citep{Barnard1919}, it is located at the  western edge of the L1630 molecular cloud. The nearby O9.5V star $\sigma$ Orionis (hereafter $\sigma$ Ori), located to the southwest of the Horsehead Nebula, emits ionizing radiation which drives the HII region, IC 434, and so sculpts the Horsehead Nebula and its PDR \citep{Abergel2003, Pound2003}. The distance to $\sigma$ Ori has been measured as 387.51 $\pm$ 1.32 pc using interferometric  parallax observations \citep{Schaefer2016}. We thus adopt 388 pc as the distance to the Horsehead Nebula. Two submillimeter sources were identified in the Horsehead Nebula using the SCUBA camera on the James Clerk Maxwell Telescope (JCMT) by \citet{WardThompson2006}. These two sources are named as B33-SMM1 and B33-SMM2, and are located in the ``head" and ``neck" of the ``horse's head'' structure, respectively. B33-SMM1 is the PDR driven by $\sigma$ Ori, while B33-SMM2 is an embedded starless core \citep{WardThompson2006}.

Magnetic fields within PDRs have been studied using polarized dust emission (e.g., \citealt{Pattle2018}). Dust grains spun up by radiative torques imparted by interactions with stellar photons process around magnetic field lines with their minor axes parallel to the magnetic field direction \citep{Lazarian2007}. Plane-of sky magnetic field orientations are thus determined by rotating the polarization angle of polarized dust emission by 90 degrees. Magnetic field strengths can be estimated from polarized dust emission measurements using the Davis-Chandrasekar-Fermi (DCF) method \citep{Davis1951, ChanFer1953}. 

By observing polarized dust emission from the Horsehead Nebula, we can study the magnetic fields in both the PDR and the starless core. The role of magnetic fields in the evolution of starless cores and PDRs is uncertain because only a few measurements of magnetic field strengths in such sources have been made (e.g., \citealt{Pattle2018}; \citealt{Karoly2020}). While magnetic fields in starless cores typically show an ordered and linear geometry (e.g., \citealt{Karoly2020}), those in PDRs are more complex. \citet{Arthur2011} predicted magnetic field geometry in PDRs using three-dimensional radiation-magnetohydrodynamic simulations. Magnetic fields in the pre-existing molecular cloud are compressed by the radiation from the massive young star in their simulations. As a result, magnetic field lines are aligned parallel to the shell-like molecular gas structure bordering the HII region, and to the boundaries of pillars  protruding from the shell into the HII region. Magnetic fields thus show hairpin structures at the ends of pillars. 

Here, we present polarized dust emission observations of the Horsehead Nebula obtained by the JCMT at 850 $\mu$m. We show plane-of-sky magnetic field orientations in the PDR and starless core (hereafter, SMM1 and SMM2, respectively). We estimate magnetic field strengths in both sources using a modified DCF method \citep{Hildebrand2009}. From the estimated magnetic field strengths, we derive mass-to-flux ratios and Alfv\'en Mach numbers in SMM1 and SMM2. We additionally estimate magnetic, gravitational and magnetic energies in SMM2 and discuss their relative importance. Based on these results, we suggest a schematic view of, and formation scenario for, the Horsehead Nebula.

This paper is organized as follows: in Section \ref{sec:observe}, we describe the JCMT observations of the Horsehead Nebula and the data reduction processes. We present the polarization angle distribution in Section \ref{sec:result}. Measurements of magnetic field strengths and other parameters such as mass-to-flux ratios, Alfv\'en Mach number and energies are presented in Section \ref{sec:discussion}. We summarize our results in Section \ref{sec:conclusions}.

\section{Data reduction} \label{sec:observe}

\begin{figure*}[htb!]
\epsscale{0.9}
\plotone{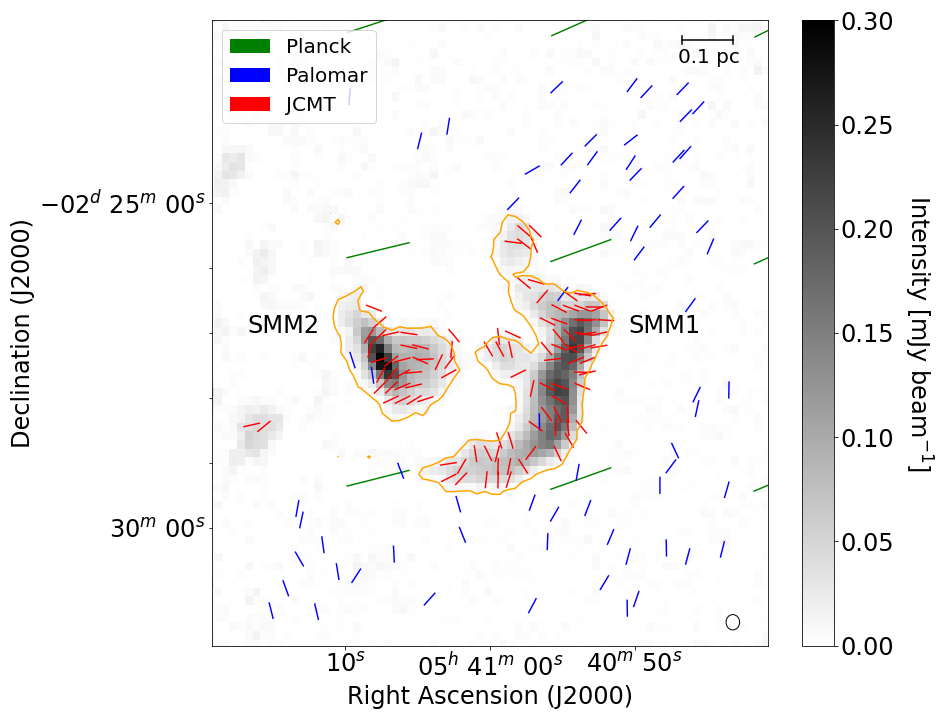}
\caption{Map of SCUBA-2/POL-2 850 $\mu$m intensity in the Horsehead Nebula. Magnetic field orientations are shown as red, blue and green segments, which were obtained using the JCMT (this work), the Palomar Observatory \citep{Zaritsky1987}, and Planck \citep{Planck2015}, respectively. The orange contour level marks a total intensity of 9 mJy beam$^{-1}$. The circle in the lower right corner shows the JCMT beam size of 14$\farcs$1 at 850 $\mu$m.
\label{fig:jcmt}}
\end{figure*}

\begin{figure*}[htb!]
\epsscale{0.9}
\plotone{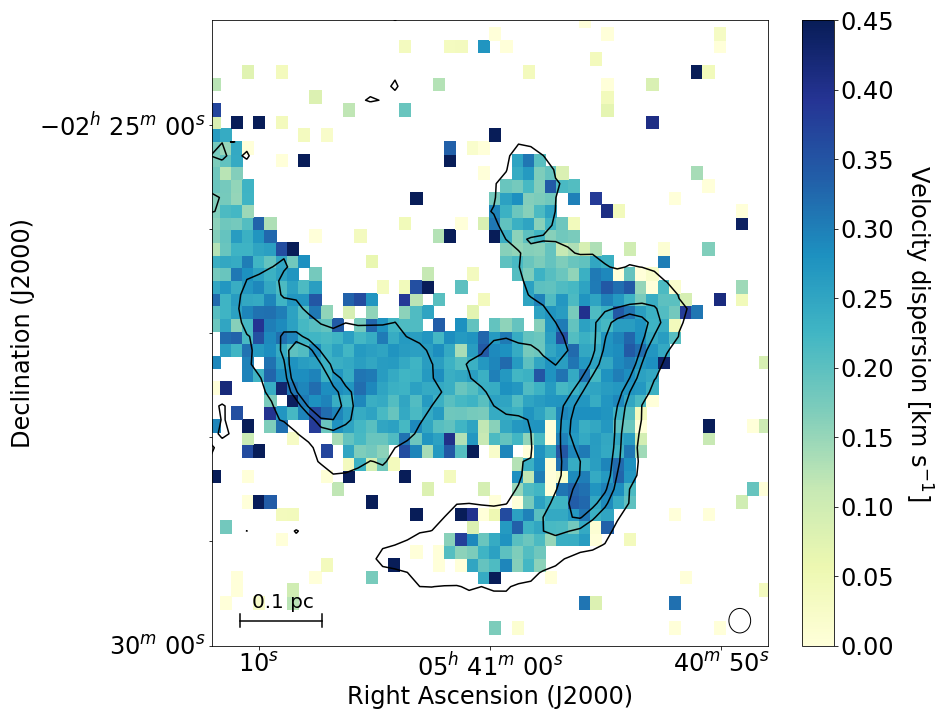}
\caption{Map of C$^{18}$O velocity dispersion in the Horsehead Nebula. The contours show SCUBA-2 850 $\mu$m flux densities of 9, 100 and 150 mJy beam$^{-1}$. The circle in the lower right-hand corner shows the JCMT beam size of 14$\farcs$1 at 850 $\mu$m.
\label{fig:c18o}}
\end{figure*}

We used observations of polarized dust emission and the C$^{18}$O $J=3\to 2$ spectral line in the Horsehead Nebula, obtained using the Submillimetre Common-User Bolometer Array 2 (SCUBA-2) camera and its POL-2 polarimeter, and with the Heterodyne Array Receiver Program (HARP) on the JCMT. The data were obtained in Director's Discretionary Time (DDT) under project code M18BD002 (PIs: Mallory Go, Harriet Parsons) as part of the Maunakea Scholars program\footnote{\url{https://maunakeascholars.com}}. We here discuss the data reduction processes for, and the properties of the reduced maps of, the two data sets. 

\subsection{SCUBA-2/POL-2 data}

Polarized dust emission from the Horsehead Nebula was observed using the POL-2 polarimeter inserted in front of the SCUBA-2 bolometer camera \citep{Holland2013} on the JCMT at 450 and 850 $\mu$m simultaneously,  although we consider only the 850 $\mu$m data in this work. The field was mapped 10 times in 32-minute observing blocks with SCUBA-2/POL-2 on 2018 November 21. The total on-source time was about 5 hours 20 minutes. The observations were conducted in JCMT Weather Band 1, in which the atmospheric opacity at 225 GHz ($\tau_{225}$) $\leq$ 0.05. The effective beam size of the JCMT is 14$\farcs$1 at 850 $\mu$m \citep{Dempsey2013}.

We reduced the data using the Submillimetre User Reduction Facility (SMURF) package \citep{Chapin2013} from the Starlink software suite \citep{Currie2014}. The $pol2map$\footnote{\url{http://starlink.eao.hawaii.edu/docs/sun258.htx/sun258ss73.html}} command is used to reduce POL-2 data. In the first step of the data reduction process, $pol2map$ is used to convert raw bolometer timestreams for each observation into separate Stokes Q, U, and I timestreams, and to create initial Stokes $I$ maps from the Stokes $I$ timestreams using the SMURF routine $makemap$ \citep{Chapin2013}. The  coadd of the initial Stokes $I$ maps is used to make masks defining areas of astrophysical emission. In the second step of the data reduction process, $pol2map$ is used to create final Stokes $I$, $Q$ and $U$ maps and a polarization segment (or half-vector) catalogue. In this second step, we used the $skyloop$ and $mapvar$ parameters in $pol2map$. $skyloop$  improves signal-to-noise ratio and image fidelity by reducing all 10 observations concurrently rather than consecutively, while $mapvar$ calculates the variances of the Stokes $I$, $Q$ and $U$ maps from the spread of pixel data values between the individual observations. We used  the August 2019\footnote{\url{https://www.eaobservatory.org/jcmt/2019/08/new-ip-models-for-pol2-data/}} instrumental polarization model to correct the Stokes $Q$ and $U$ maps. A detailed description of the the POL-2 data reduction process is given in, e.g., \citet{Pattle2021}; \citet{Konyves2021}. In the second step of the data reduction process, we used a pixel size of 8$''$ instead of the default 4$''$ pixel size to increase the signal-to-noise ratio (SNR) of the final Stokes $I$, $Q$ and $U$ maps, but used masks defined using the 4$''$-pixel-size Stokes $I$ map from step 1, in order to provide sufficient constraints on the mapmaker. A detailed investigation into the effect of pixel size on POL-2 data reduction and the choice to use 8$''$ pixels is presented by Karoly et al. (in prep.). 

The native units of the reduced Stokes $I$, $Q$ and $U$ maps are pW. We converted the maps to Jy beam$^{-1}$ using a Flux Conversion Factor (FCF) of 495 Jy beam$^{-1}$ pW$^{-1}$ \citep{Mairs2021}, multiplied by a factor of 1.35 to account for additional flux losses in POL-2 \citep{Friberg2016}. As we used 8$''$ rather than 4$''$ pixels, we further multiplied the maps by a factor of 1.12, determined from SCUBA-2 calibration data (Karoly et al. in prep.).  Figure \ref{fig:jcmt} shows an image of the 850 $\mu$m total intensity (Stokes $I$) map of the Horsehead Nebula in units of Jy beam$^{-1}$, on which polarization segments are overplotted.
 
 We binned the polarization segments obtained using the data reduction processes described above to a pixel size of 12$''$, which is the JCMT 850 $\mu$m primary beam size, and is similar to the JCMT 850 $\mu$m effective beam size of $14\farcs1$. The polarization fractions $p$ listed in the polarization half-vector catalogues are debiased, and are given by
 \begin{equation}
     p = \frac{(Q^2 + U^2 - 0.5[(\delta Q)^2+(\delta U)^2])^{1/2}}{I},
 \end{equation} where $\delta Q$ and $\delta U$ are the square roots of the measured variances on $Q$ and $U$ respectively. The polarization angles $\theta$ are given by
 \begin{equation}
     \theta=\frac{1}{2}\arctan\left(\frac{U}{Q}\right) \times \frac{180^\circ}{\pi}.
 \end{equation} Uncertainties on $p$ and $\theta$ are given by
 \begin{equation}
     \delta p = \sqrt{\frac{(Q^2\delta Q^2+U^2\delta U^2)}{I^2(Q^2+U^2)}+\frac{\delta I^2(Q^2+U^2)}{I^4}},
 \end{equation}
 and
 \begin{equation}
     \delta \theta =\frac{1}{2}\frac{\sqrt{(Q^2\delta U^2+U^2\delta Q^2)}}{(Q^2+U^2)} \times \frac{180^\circ}{\pi},
 \end{equation}
 respectively. We rotated these polarization angles by 90 degrees to show magnetic field orientations in the Horsehead Nebula in Figure \ref{fig:hst} and Figure \ref{fig:jcmt}, and in subsequent analysis of magnetic field direction. The maps and half-vector catalogs used in this work are available at [DOI to be inserted in proof].

\subsection{HARP data}

Three CO isotopologues, $^{12}$CO ($J = 3-2$), $^{13}$CO ($J = 3-2$) and C$^{18}$O ($J = 3-2$), were observed in the Horsehead Nebula under project code M18BD002 using HARP. The $^{13}$CO and C$^{18}$O observations were made simulatenously on 2018 December 30. All observations of the CO isotopologues were carried out in weather band 2 (0.05 $<$ $\tau_{225}$ $\leq$ 0.08). In this work, we use only the C$^{18}$O data, which like the other isotopologues has a critical density of $\sim 10^{4}$ cm$^{-3}$, and which, with the lowest fractional abundance of the three isotopologues, is able to trace gas density to the highest optical depth \citep[e.g.][]{Rigby2016}.  With a rest frequency of 329.331 GHz $\approx 911$ $\mu$m, the C$^{18}$O data has comparable resolution to the SCUBA-2 850 $\mu$m data.

To estimate velocity dispersion values in the Horsehead Nebula, we reduced  the C$^{18}$O data using the ORAC Data Reduction (ORAC-DR) pipeline and the Kernel Application Package (KAPPA; \citealt{Currie2008}) in the Starlink software suite \citep{Jenness2013}. We obtained a reduced C$^{18}$O data cube in which the pixel size, beam size and spectral resolution are 7$''$, 15.3$''$ and 0.05 km s$^{-1}$, respectively. To improve the SNR, we smoothed the data cube to a spectral resolution of 0.15 km s$^{-1}$. Figure \ref{fig:c18o} shows the velocity dispersion of  the C$^{18}$O data, which was obtained by fitting the C$^{18}$O spectra with a single Gaussian profile. The contours in the figure show total intensities at 850 $\mu$m of 9, 100 and 150 mJy beam$^{-1}$. We analyzed the two regions within the lowest contour level, in which velocity dispersions of C$^{18}$O are similar to each other.

\section{Results} \label{sec:result}

\begin{figure*}[thb!]
\epsscale{0.8}
\plotone{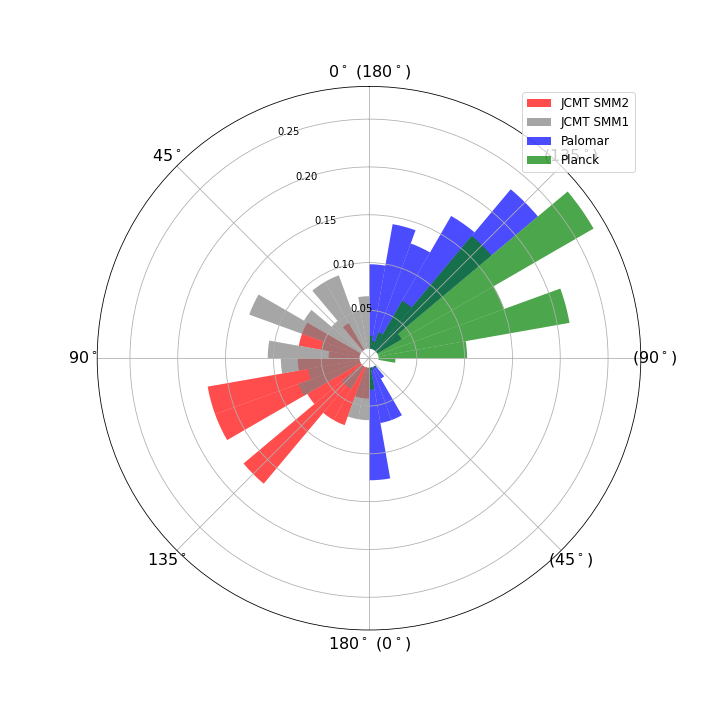}
\caption{Magnetic field orientations obtained using the JCMT (this work), Planck \citep{Planck2015}, and Palomar \citep{Zaritsky1987}. Gray and red  bars show magnetic field orientations obtained using the JCMT in SMM1 and SMM2, respectively. The magnetic field orientations obtained in SMM2 are approximately parallel to those obtained using Planck. Overall, the magnetic field orientations obtained using Palomar are shifted about 20 degrees compared to those obtained using Planck. The JCMT POL-2 histograms are shown over the angle range $0^{\circ}-180^{\circ}$ E of N, while the Planck and Palomar histograms is shown over the range $180^{\circ}-360^{\circ}$. Due to the $\pm 180^{\circ}$ ambiguity on polarization vector measurements, this range is identical to the range $0^{\circ}-180^{\circ}$, and so opposite angles on the plot agree. In Cartesian space, the area of each histogram is normalized to 1; note that the projection onto a circle means that their areas are distorted. 
\label{fig:mag_ori}}
\end{figure*}

Figure \ref{fig:hst} shows a near-infrared image of the Horsehead Nebula obtained using the HST, with magnetic field orientations overlaid. The image shown in Figure \ref{fig:hst} is a composite of two archive images obtained by the WFC3 infrared camera on the Hubble Space Telescope (HST)  in the broad band F110W (YJ) and F160W (H) filters, centered at 1153.4 and 1536.9 nm, respectively \footnote{\url{https://archive.stsci.edu/prepds/heritage/horsehead/}}.

Submillimeter observations, such as our SCUBA-2/POL-2 observations, allow us to see within the Horsehead Nebula, and show us that the Horsehead contains two dense sources: the western PDR (SMM1), and the eastern starless core (SMM2), which are labeled on Figure \ref{fig:jcmt}. The PDR and the starless core are in the ``head" and ``neck" of the ``horse's head", respectively.

Magnetic field orientations in the Horsehead Nebula are shown in Figure \ref{fig:jcmt}. As well as our JCMT data, we show magnetic field measurements made using Planck \citep{Planck2015} and the Palomar Observatory \citep{Zaritsky1987}. The Palomar polarization observations  were conducted at optical wavelengths using the four-shooter polarizing filter wheel in front of a CCD camera on the 5 m telescope of the Palomar Observatory \citep{Zaritsky1987}. Polarized starlight results from the preferential extinction of light from background stars by dust grains aligned with their major axis perpendicular to the local magnetic field direction \citep{Hiltner1949}, and so optical polarization segments in  Figures \ref{fig:hst} and \ref{fig:jcmt} directly show magnetic field orientations in the plane of the sky. The Planck observations trace polarized dust emission, and were carried out at 353 GHz, comparable to the JCMT observations. However, the full width at half maximum (FWHM) of the Planck beam is 10$'$, which is much larger than that of the JCMT at 850 $\mu$m. The polarization segments obtained by Planck and the JCMT are perpendicular to the plane-of-sky magnetic field direction, and so we rotated these segments by 90 degrees in Figures \ref{fig:hst} and \ref{fig:jcmt} to show magnetic field orientations.

\subsection{Magnetic field morphology}

Both SMM1 and SMM2 have ordered magnetic fields, but the fields in the two regions have very different magnetic field morphologies. We select those polarization segments for which $p$/$\delta p \geq 2$ and $p< 20\%$ for analysis.
Figure \ref{fig:mag_ori} shows magnetic field orientations in the Horsehead Nebula as a polar bar chart. The left-hand side of the plot shows magnetic field orientations in SMM1 and SMM2 obtained by the JCMT, while the right-hand side shows magnetic field orientations in the vicinity of the Horsehead Nebula obtained by Palomar and Planck, such that diametrically opposite angles on the plot agree. The magnetic field orientations in SMM2 have a peak at 115 $\pm$ 30 degrees, approximately perpendicular to the major axis of the source, and are broadly aligned with those measured with Planck, and similar to those measured at the Palomar Observatory.  The magnetic field orientations in SMM1 are roughly perpendicular to the curved structure of SMM1 in the plane of the sky. The magnetic field orientations in SMM1 thus occupy a wide range of angles in Figure \ref{fig:mag_ori}.

\citet{Pattle2018} presented magnetic field orientations within the photoionized pillars known as the `Pillars of Creation' in M16 at a distance of $\sim 1.8$ kpc. They found that magnetic fields are aligned along the length of the pillars, but were unable to resolve the magnetic fields in the PDRs at the pillars' tips. Our observations of the Horsehead nebula are a factor $\sim 5$ higher in linear resolution than the M16 observations, and the PDR of the Horsehead appears to be more extended in the plane of the sky than those of the pillars in M16, allowing us to resolve the magnetic field within the PDR. The magnetic field in SMM1 is perpendicular to the major axis of the PDR structure. This is quite dissimilar to POL-2 observations of the magnetic field in the Orion Bar in OMC-1, at a comparable distance to the Horsehead, in which the magnetic field runs along the length of the PDR \citep{WardThompson2017, Pattle2017}. A possible cause of the morphology which we observe in SMM1 is that we are seeing a magnetic field in the PDR which is folded back on itself along the line of sight. A detailed schematic view of the magnetic field morphology of SMM1 is presented in Section \ref{sec:view}. 

\subsection{Column density}

We calculated the column density of molecular hydrogen in SMM1 and SMM2 using \citep{Hildebrand1983}
\begin{equation}
    N(\text{H}_2) =\frac{I_{\nu}}{\mu m_{\text{H}}  \kappa(\nu) B_\nu(T)},
    \label{eq:coldensity}
\end{equation} 
where $I_{\nu}$ is the total intensity at frequency $\nu$, $\mu$ is the mean molecular weight, $m_{\text{H}}$ is the mass of a hydrogen atom, $\kappa(\nu)$ is the dust opacity and $B_\nu(T)$ is the Planck function for a dust temperature $T$. We took $\mu$=2.8 \citep{Kauffmann2008}. The dust opacity is estimated using $\kappa(\nu)=\kappa_{\nu_0}(\nu/\nu_0)^\beta$ where $\kappa_{\nu_0}$ is the dust opacity at the reference frequency $\nu_0$, and $\beta$ is the dust opacity spectral index. We took $\kappa_{\nu_0}$ = 0.1 cm$^2$ g$^{-1}$ at $\nu_0$ = 1000 GHz, and $\beta$ = 2, thus assuming a dust-to-gas mass ratio of 1:100 (\citealt{Beckwith1990}; \citealt{MotteAndr2001}; \citealt{Andre2010}). 

The dust temperatures in SMM1 and SMM2 are 22 and 15 K, respectively \citep{WardThompson2006}. We estimated mean column densities for SMM1 and SMM2 by substituting the mean Stokes I intensities for SMM1 and SMM2 of 71 and 69 mJy beam$^{-1}$ at 850 $\mu$m, as measured from our observations, into equation (\ref{eq:coldensity}). The mean Stokes $I$ intensities are estimated by averaging intensities within the orange contours in Figure \ref{fig:jcmt}, in which the left contour outlines SMM2 and the right contour outlines SMM1.
The mean column densities estimated in SMM1 and SMM2 are 3.9 $\times 10^{21}$ and 6.7 $\times 10^{21}$ cm$^{-2}$. The peak column densities of SMM1 and SMM2 are 1.3 $\times 10^{21}$ and 3.0 $\times 10^{22}$ cm$^{-2}$, which are comparable to within a factor of a few with those estimated by \citet{WardThompson2006}.

\begin{figure*}[htb!]
\epsscale{1.0}
\plotone{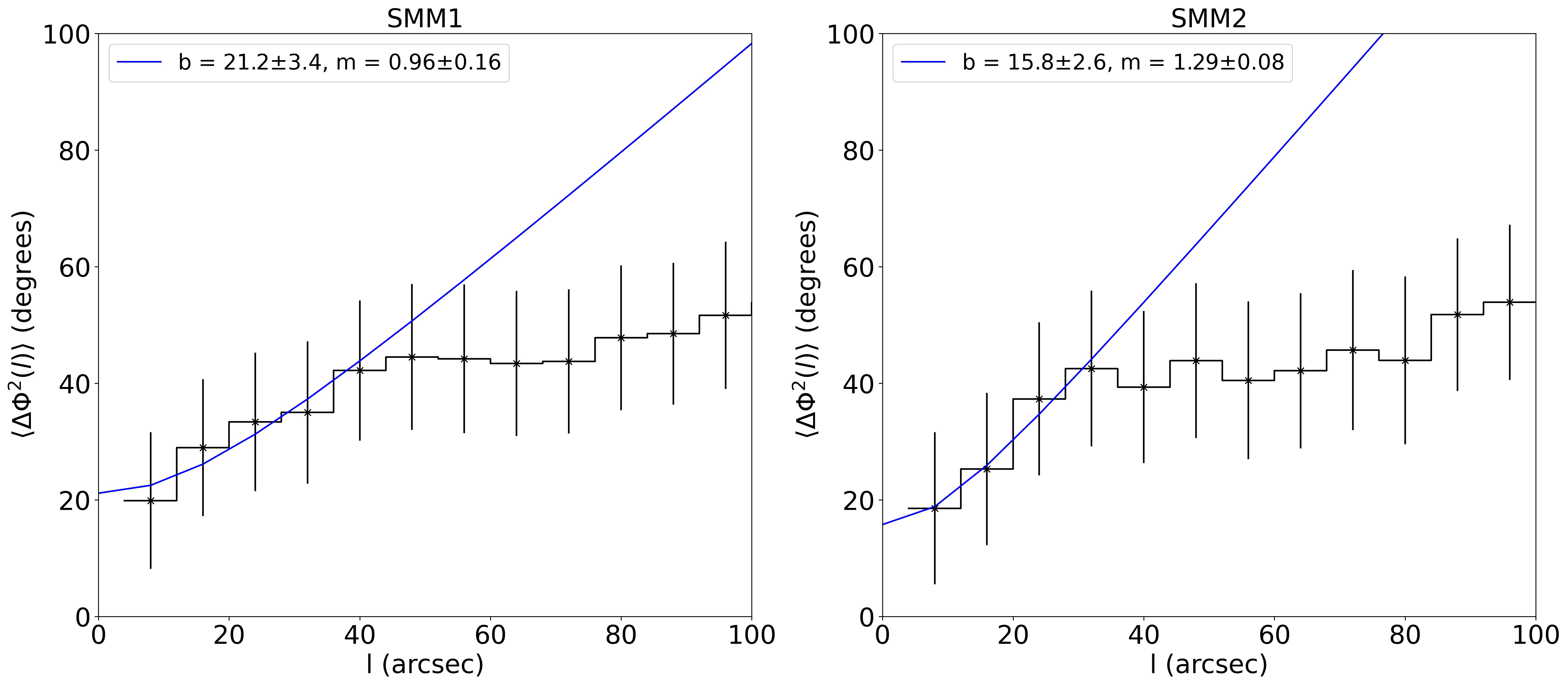}
\caption{Angular dispersion functions measured in SMM1 (left) and SMM2 (right). Blue lines show the quadratic function fitted to the angular dispersion function. The fitted $b$ and $m$ values are shown in left upper corners of each panel. Note that $b$ is here given in degrees for ease of interpretation.
\label{fig:sf}}
\end{figure*}

\subsection{Magnetic field strengths} \label{subsec:B}

We estimated magnetic field  strengths in the Horsehead Nebula using the angular dispersion function \citep{Hildebrand2009} implementation of the DCF method \citep{Davis1951, ChanFer1953}. The DCF method is widely used to estimate magnetic field strengths in star-forming regions from dust polarization observations \citep[e.g.][and refs. therein]{Pattle2022}. DCF estimates magnetic field strength from the gas density, velocity dispersion and polarization angle dispersion in a given region. The fundamental assumption of DCF is that the distortion of magnetic field lines by small-scale turbulent gas motions is indicative of the Alfv\'en Mach number of the gas, and can be estimated as a polarization angle dispersion in the region. In traditional implementations of DCF, polarization angle dispersion is measured under the assumption of a uniform underlying magnetic field direction. However, magnetic field lines can show ordered variation and curvature due to gravitational collapse, rotation, shocks or outflows. In order to estimate magnetic field strengths using traditional DCF, it is thus essential to assume a model for the structure of the mean magnetic field.

\citet{Hildebrand2009} suggested a method to separate the ordered and turbulent components of the magnetic field, a structure-function based approach to DCF known as the angular dispersion function. To calculate the angular dispersion function, we calculate the differences in polarization angles between individual pairs of polarization segments, $\Delta \theta (l) \equiv \theta(x) - \theta (x+l)$, where $\theta (x)$ is the polarization angle of a segment at position $x$, and $\theta (x+l)$ is the polarization angle of a segment separated from $x$ by a distance $l$. If the number of pairs is given by $N(l)$, the angular dispersion function is then given by
\begin{equation}
\langle\Delta \theta^2 (l)\rangle^{1/2} \equiv \left[ \frac{1}{N(l)}\sum^{N(l)}_{i=1}\Delta \theta (l)^2 \right] ^{1/2}.
\end{equation}
 The angular dispersion function can, at small $l$, be fitted with the quadratic function
 \begin{equation}
     \langle\Delta \theta ^2 (l)\rangle = b^2 + m^2l^2 +\sigma_M^2(l),
\label{eq:dispersion}
 \end{equation}
 where $ml$ is the mean magnetic field (referred to as the large-scale field by \citet{Hildebrand2009}), $b$ is the turbulent dispersion about the mean magnetic field and $\sigma_M(l)$ is the measurement uncertainty. \citet{Hildebrand2009} expressed polarization angle dispersion as the ratio of the turbulent to large-scale magnetic field strength, which is given $b/\sqrt{2-b^2}$. When the turbulent component of the magnetic field is much smaller than the ordered component, i.e. $b\ll1$ rad, $b/\sqrt{2-b^2} \to b/\sqrt{2}$. In this case, the DCF equation for magnetic strength can be expressed as
\begin{equation} 
B_{pos} = \sqrt{8\pi \rho}\frac{\sigma_v}{b},
\label{eq:dcf}
\end{equation}
where $B_{pos}$ is the magnetic field strength in the plane of the sky, $\rho$ is the mass density of gas coupled to the magnetic field 
and $\sigma_v$ is the velocity dispersion of the gas. The mass density is given by $\rho=\mu m_{\text{H}} n({\text{H}}_2)$, where $\mu$ is the mean molecular weight, taken to be 2.8 \citep{Crutcher2004a, Kauffmann2008}, $m_{\text{H}}$ is the mass of a hydrogen atom, and $n({\text{H}}_2)$ is the volume number density of the gas. 

We estimated each of the parameters in equation (\ref{eq:dcf}) in order to measure magnetic field strengths in SMM1 and SMM2. We fitted the quadratic model of the angular dispersion function, as given in equation (\ref{eq:dispersion}), to the dispersion of polarization angle differences in SMM1 and SMM2 as a function of the distance between pairs of polarization segments, as shown in Figure~\ref{fig:sf}. We fitted the data in the range $l<36''$. The fitted $b$ values are $21.2^{\circ}\pm3.4^{\circ}$ and $15.8^{\circ}\pm2.6^{\circ}$ in SMM1 and SMM2, respectively. Note that the units of $b$ in equation~(\ref{eq:dcf}) are radians.

We estimated volume density values from the column density values obtained in Section \ref{sec:result}. We assumed effective radii for SMM1 and SMM2 of 0.15 and 0.10 pc, which we took to be the radii of circles with areas equal to the areas within the orange contours shown in Figure \ref{fig:jcmt}. The volume density is estimated using
\begin{equation}
n(\text{H}_2)\times\frac{4}{3}\pi r^3 = N(\text{H}_2)\times\pi r^2,
\end{equation}
where $n(\text{H}_2)$ is the volume density and $r$ is the effective radius. We thus estimate mean volume densities in SMM1 and SMM2 of 6.4$\times$10$^3$ and 1.7$\times$10$^4$ cm$^{-3}$.

 We estimated non-thermal velocity dispersion values from our C$^{18}$O spectral line data. We fitted each pixel of the observed line data with a single Gaussian profile. The non-thermal velocity dispersion is given by
\begin{equation}
    \sigma_v^2=\sigma_{obs}^2-\frac{kT_k}{m_{\text{C}^{18}\text{O}}},
\label{eq:vel}
\end{equation}
where $\sigma_{obs}$ is the estimated dispersion of the Gaussian profile (Figure \ref{fig:c18o}), $k$ is the Boltzmann constant, $T_k$ is the kinetic temperature and $m_{ \text{C}^{18} \text{O}}$ is the mass of C$^{18}$O molecule. We took $T_k$ to be equal to the dust temperatures of 22 and 15 K in SMM1 and SMM2, respectively.
The mean observed velocity dispersions in SMM1 and SMM2 are 0.11 and 0.10 km~s$^{-1}$. These dispersions are dominated by non-thermal motions; the non-thermal velocity dispersions given by equation~(\ref{eq:vel}) agree with these values to two decimal places.

By substituting the above values into equation (\ref{eq:dcf}), we estimated magnetic field strengths of 56$\pm$9 and 129$\pm$21 $\mu$G in SMM1 and SMM2, respectively. The uncertainties on these magnetic field strengths were obtained by propagating the fitting uncertainties on $b$ for the two sources. 

\section{Discussion} \label{sec:discussion}

In this paper we present the first estimates of magnetic field strengths in the Horsehead Nebula. There are several previous works which have estimated magnetic field strengths in starless cores and a handful of estimates of magnetic field strengths in PDRs at submillimeter wavelengths. \citet{Pattle2018} estimated a magnetic field strength in a pillar in the Eagle Nebula (M16) of 170--320 $\mu$G, larger than that which we estimate SMM1. There are a number of differences between the two regions: the volume density of the pillar in M16 is an order of magnitude larger than that in SMM1, and the HII regions in M16 are driven by a  massive star cluster (NGC 6611), while that in the Horsehead Nebula is formed by a single O star, suggesting a more simple star formation than that in M16. Moreover, the magnetic field strength estimated for M16 is that of the pillar, not of that in the PDR at the pillar's head. All of these factors could result in the different magnetic field strengths estimated in SMM1 and M16. However, DCF measurements derived from SOFIA HAWC+ far-infrared observations of the Orion Bar  also show magnetic field strengths of up to a few hundred $\mu$G \citep{Guerra2021}, more similar to that of M16 than that of the Horsehead.  While the Orion Bar is an extended PDR at a similar distance to the Horsehead, it is again driven by a stellar cluster (the Trapezium Cluster) rather than by a single star.

There are a relatively small number of estimates of magnetic field strengths in starless cores, due to their low brightness. 
Magnetic field strengths in the starless cores L183,  $\rho$ Ophiuchus C, L1689B and SMM16 range from 72 to 284 $\mu$G \citep{Liu2019, Karoly2020, Pattle2021}. These values are comparable to the magnetic field strength which we estimate in SMM2.

\subsection{The Energetic Importance of Magnetic Fields in the Horsehead Nebula}

\begin{figure*}[htb!]
\epsscale{0.8}
\plotone{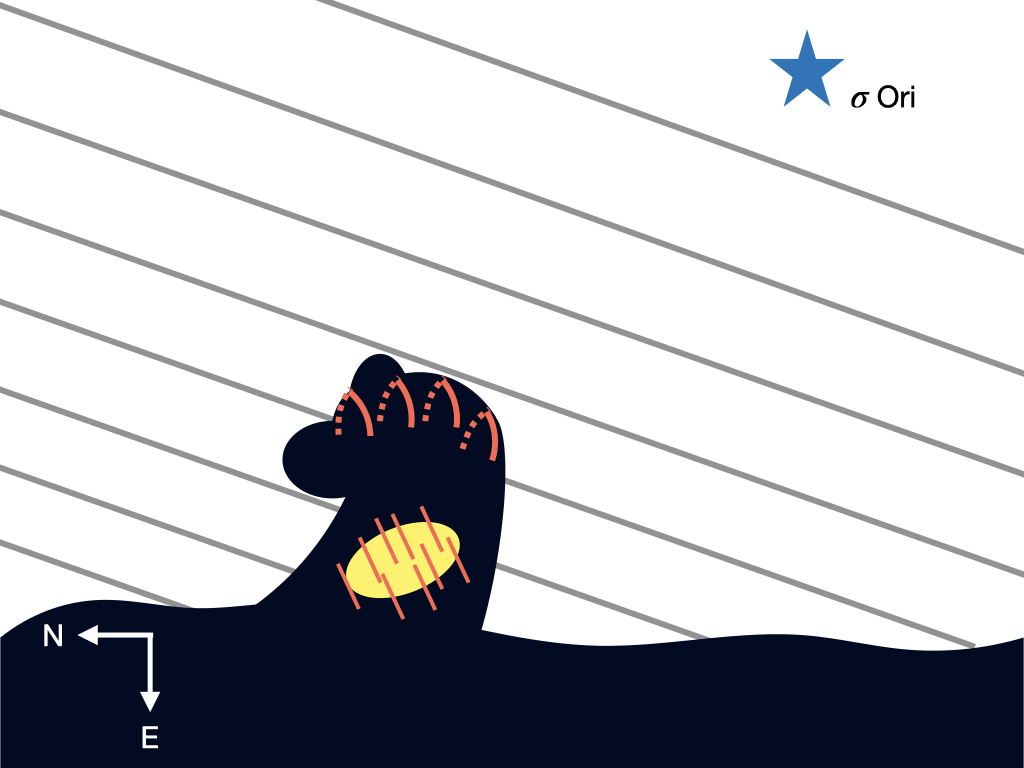}
\caption{A schematic view of the Horsehead Nebula. The Horsehead Nebula is shown in the lower left of the figure. The starless core SMM2 is shown as a yellow ellipse. The red lines show magnetic field lines within SMM1 and SMM2. The gray lines show magnetic field lines in the HII region. The blue star is $\sigma$ Ori, which is located behind and to the southwest of the Horsehead Nebula. The line of sight to the observer is perpendicular to the page, or slightly inclined  therefrom.
\label{fig:view}}
\end{figure*}

The relative energetic importance of magnetic fields in the Horsehead Nebula can be estimated using the mass-to-flux ratio and Alfv\'en Mach number to compare magnetic fields to self-gravity and internal turbulent motions. We derive mass-to-flux ratios and Alfv\'en Mach numbers for both SMM1 and SMM2. Additionally, we estimate the magnetic, gravitational and turbulent energies in SMM2, and consider their relative importance. From these results, we discuss the role of magnetic fields in the evolution of the Horsehead Nebula. 

\subsubsection{Mass-to-flux ratio}
\label{subsubsec:massflux}

The mass-to-flux ratio is widely used to study the relative importance of magnetic fields and self-gravity in molecular clouds and the structures within them (e.g., \citealt{Mouschovias1976}; \citealt{Crutcher2004a}). The mass-to-flux ratio is typically expressed as $\lambda$, the ratio of the measured mass-to-flux  ratio to its critical value. When we use the critical value for a magnetized disk supported by magnetic field against gravity, $1/2\pi\sqrt{G}$ \citep{Nakano1978}, $\lambda$ can be parameterized as,
\begin{equation}   
\lambda=7.6\times10^{-21}\frac{N(\text{H}_2)}{B},
\end{equation}
\citep{Crutcher2004a}, where $B$ is the three-dimensional magnetic field strength in units of $\mu$G, and $N(\text{H}_2)$ is in units of cm$^{-2}$. If $\lambda < 1$, the structure under consideration is supported by magnetic fields against gravitational collapse, and is referred to as being `magnetically sub-critical'.  If $\lambda > 1$, the structure's internal magnetic field cannot prevent gravitational collapse, and so the structure is referred to as being `magnetically super-critical'.

If we assume $B$ $\sim$ $B_{pos}$, the mass-to-flux ratios in SMM1 and SMM2 are 0.5$\pm$0.1 and 0.4$\pm$0.1, respectively. \citet{Crutcher2004b} noted that on average, three-dimensional magnetic field strength is related to the magnetic field strength in the plane of the sky by $B = 4/\pi \times B_{pos}$. While it should be noted that this strictly holds only for a statistical ensemble of measurements, the mass-to-flux ratios obtained using this estimate of $B$ are  0.41$\pm$0.07 in SMM1 and 0.31$\pm$0.06 in SMM2.

 A long-standing issue in DCF studies is that the measured polarization angle dispersion is integrated over some number $N$ turbulent eddies along the line of sight, and so the magnetic field strength will be overestimated by a factor $\sqrt{N}$ \citep[e.g.][and refs. therein]{Pattle2019}. \citet{Cho2016} proposed a method to estimate $\sqrt{N}$ from the ratio of the mean velocity dispersion (linewidth) to the dispersion of centroid velocities of the spectral line being used for the DCF analysis. We calculated this ratio for our C$^{18}$O measurements in SMM1 and SMM2. From the ratio, we estimated that there are 1 and 4 turbulent eddies along the line of sight in SMM1 and SMM2, respectively. The magnetic field strength in SMM2 may thus be overestimated by a factor of $\sqrt{4} = 2$. If we include this correction factor, the estimated three-dimensional mass-to-flux ratio in SMM2 becomes 0.6$\pm$0.1. 

Irrespective of whether or not we attempt to account for the three-dimensional magnetic field strength or the number of eddies, the mass-to-flux ratios which we infer are less than unity in both SMM1 and SMM2. This suggests that the magnetic fields in SMM1 and SMM2 are strong enough to support them both against gravitational collapse. This is not a surprising result in SMM1, where we expect gas and radiation pressure from the HII region to play a more significant role than self-gravity in the evolution of the PDR. The starless core SMM2 appears closer to magnetic criticality than SMM1, although neither SMM1 nor SMM2 have mass-to-flux ratios very different from unity.

\subsubsection{Alfv\'en Mach number}

The Alfv\'en Mach number ($M_A$) is used to parameterize the relative importance of non-thermal motions -- in this case, internal small-scale turbulence -- and magnetic fields. $M_A$ is defined as the ratio of non-thermal (turbulent) velocity dispersion to Alfv\'en velocity \citep[e.g.][]{Crutcher1999}, and is equivalent to the ratio $b/\sqrt{2-b^2}$ discussed in section \ref{subsec:B}. We thus estimate Alfv\'en Mach numbers in SMM1 and SMM2 of 0.27$\pm$0.05 and 0.20$\pm$0.03, respectively, where uncertainties are again estimated by propagating the uncertainty on the fitted value of $b$.  If the Alfv\'en Mach number is less than unity, magnetic pressure dominates over internal turbulent pressure. If we apply our $\sqrt{N}=2$ \citet{Cho2016} correction factor to SMM2, its Alfv\'enic Mach number becomes 0.40$\pm$0.06. These results suggest that magnetic pressure exceeds internal turbulent pressure in both SMM1 and SMM2. However, SMM1 is also affected by significant external pressure from the nearby O star and its associated HII region \citep{Habart2005}.

\subsubsection{Energy balance in SMM2}

The magnetic field in the starless core SMM2 is ordered, linear and approximately parallel to the minor axis of the core.  This is similar to magnetic field geometries observed in many other low-mass starless cores \citep[e.g.][]{WardThompson2000,Kirk2006,Liu2019,Coude2019,Pattle2021}.  Such a geometry is suggestive of a dynamically important magnetic field which has imposed a preferred direction on the collapse of the core \citep{Mouschovias1976}.  Again in common with other starless and prestellar cores, we see no clear sign of the `hourglass' field geometry which would be indicative of the field being dragged in by gravitational collapse \citep{Fiedler1993}.  This suggests that the core is likely to be stable, or in the earliest stages of collapse.

\citet{WardThompson2006} suggested that SMM2 is an embedded core which existed before the formation of the Horsehead PDR. The magnetic field which we observe is consistent with this interpretation, being ordered, linear and consistent with both the average Planck magnetic field direction in the region -- likely that of the parent molecular cloud -- and that observed in the HII region by the Palomar Observatory. Simulations of the expansion of magnetized HII regions suggest that magnetic field orientation is largely unchanged by the free passage of a plane-parallel shock front \citep{Henney2009}, and so we can treat the magnetic field traced by the Palomar observations in the HII region as indicative of the magnetic field direction in the pre-shock molecular gas. This suggests that the magnetic field in SMM2 is inherited from its parent cloud and not the result of interaction with the HII region. We thus consider the energetic balance of the core under the assumption that the core is embedded within molecular gas, and is possibly further sheltered by the SMM1 PDR \citep{WardThompson2006}, and therefore is not in direct interaction with the HII region. 
However, we note that in the diffuse western part of SMM2, the magnetic field is broadly parallel to the major axis of the SMM1 PDR and perpendicular to the direction of incident radiation from $\sigma$ Ori, suggesting either that the periphery of the core may be interacting with the HII region, or that this emission comes from a different, hotter gas component along the same line of sight.

We estimated kinetic, gravitational and magnetic energies in SMM2 using the relations
\begin{align}
    E_K &=\frac{3}{2}M\sigma_{\text{tot}}^2, \\
    E_G &=-\frac{3}{5}\frac{GM^2}{R}, \\
    E_B &=\frac{1}{2}MV_A^2,
\end{align}
where $E_K$ is kinetic energy, $M$ is the mass of SMM2, $\sigma_{\text{tot}}$ is the observed velocity dispersion of C$^{18}$O (thus accounting for both thermal and non-thermal kinetic energy), $E_G$ is the gravitational potential energy, $G$ is the gravitational constant, $R$ is the effective radius of SMM2 defined in Section~\ref{subsec:B}, $E_B$ is the magnetic energy,  and $V_A$ is the Alfv\'en velocity, $B/\sqrt{4\pi\rho}$.   The mass of SMM2 is given by
\begin{equation}
M = \sum_{i=1}^N \frac{I_{850}A}{\kappa_{\nu(850)} B_{\nu(850)} (T)},
\end{equation}
where $N$ is the number of pixels within SMM2, $I_{850}$ is the per-pixel intensity at 850 $\mu$m, $A$ is the pixel area, $\kappa_{\nu(850)}$ and $B_{\nu(850)}(T)$ are the dust opacity and the Planck function at dust temperature $T$ at 850 $\mu$m, respectively.  We determined a mass of 5.1 M$_\odot$ for SMM2, comparable to the $\sim 4$ M$_\odot$ given for the source by \citet{WardThompson2006}. 

We estimated kinetic, gravitational potential and magnetic energies of 8.0, $-8.8$ and 68 $\times$ 10$^{42}$ erg in SMM2, respectively. The kinetic and gravitational energies which we estimate in SMM2 are thus comparable to one another, but an order of magnitude smaller than our estimated magnetic energy. These kinetic and gravitational potential energies are comparable to those estimated for SMM2 by \citet{WardThompson2006}, of $\sim 14$ and $\sim -21\times 10^{42}$ erg respectively (note that these values are corrected for the slightly different distance which they assume).

In the above calculation, we used our estimated plane-of-sky magnetic field strength, $B_{pos}$, to estimate the magnetic energy. If we instead use our estimate of total magnetic field strength, $B = 4/\pi \times B_{pos}$, the estimated magnetic energy becomes 110 $\times$ 10$^{42}$ erg, approaching two orders of magnitude larger than the other terms. However, if we also consider the correction for the number of turbulent eddies along the line of sight in SMM2 which we discuss in Section~\ref{subsubsec:massflux}, our estimate of total magnetic energy becomes 27 $\times$ 10$^{42}$ erg. This value is only three times larger than our estimated gravitational potential and internal kinetic energies, and comparable to, although still larger than, the \citet{WardThompson2006} values.

Despite the large magnetic energy which we estimate, SMM2 shows no indication of expanding under internal magnetic pressure, or of otherwise being significantly out of virial equilibrium. We propose two non-mutually exclusive hypotheses to explain the energy balance which we estimate in SMM2. The first is that magnetic field strengths obtained using the DCF method are likely to be systematically overestimated, typically by a factor  $\sim 3-5$ \citep{Pattle2022}. This could cause a potentially significant overestimation of magnetic energy. However, our inclusion of the \citet{Cho2016} correction factor likely mitigates against this effect \citep{Liu2021}.  Our second hypothesis is that the starless core is at least partially confined by the weight of the surrounding molecular cloud, as external gas pressure can contribute significantly to the energy balance of starless cores \citep[e.g.][]{Pattle2015}. Alternatively, if SMM2 is not sheltered from the HII region, photon pressure may play a significant role in its confinement \citep{WardThompson2006}.

\subsection{Scenario to form the Horsehead Nebula PDR magnetic field geometry} \label{sec:view}

 Broadly, both simulations \citep[e.g.][]{Arthur2011} and previous observations suggest that magnetic fields in molecular gas typically run plane-parallel to the interface with an HII region, whether in a PDR \citep{WardThompson2017}, or in columns sheltered behind PDRs \citep{Pattle2018}.  In the heads of pillars, this is predicted to result in `hairpin' magnetic field structures \citep[e.g.][]{Arthur2011}, in which the pre-shock magnetic field is bent back on itself as the pillar forms from its parent molecular cloud.

The SMM1 PDR appears to be a relatively extended, bar-like PDR sitting at the head of a broad and short pillar.  The plane-of-sky magnetic field morphology which we observe is broadly perpendicular to the projected major axis of the PDR.  Such a projected geometry could result from a plane-parallel magnetic field if either (1) we observe hairpin magnetic field lines, with the pre-shock magnetic field lines having been folded back on themselves along the line of sight, and/or (2) the Horsehead pillar is slightly inclined with respect to the plane of the sky, and we observe in projection magnetic fields which are running parallel to the plane of the SMM1 PDR, which is itself extended along the line of sight.  Although we appear to be observing the $\sigma$ Ori-Horsehead interaction nearly edge-on \citep{Abergel2003}, $\sigma$ Ori may be located slightly behind the Horsehead \citep{Pound2003}, which may lend weight to the latter of these scenarios. In both of these cases, we would expect the average orientation of the magnetic field to be inherited from that in the pre-shock molecular cloud, and so it is not surprising that the average magnetic field in the densest parts of SMM1 is similar to that in the apparently less disturbed SMM2.

Figure \ref{fig:view} shows a schematic view of the Horsehead Nebula which illustrates our proposed magnetic field geometry. The gray lines show magnetic field lines in the HII region, which likely trace the initial (pre-shock) magnetic field direction, while the magnetic field lines in SMM1 and SMM2 are shown as red lines. The O star, $\sigma$ Ori, is located to the southwest of the pillar. The observer views the region either edge-on, or slightly inclined along the line of sight. The average field direction in the HII region is 161 degrees east of north, the mean of the magnetic field orientation angles obtained from Palomar observations. In SMM1, our proposed hairpin magnetic field geometry is shown: the field lines are aligned along the pillar, and folded or curved back on themselves in the PDR at the head of the pillar (e.g., \citealt{Arthur2011, Pattle2018}). The magnetic field in SMM2 is linear, with a mean field direction of 115 degrees east of north, as estimated from JCMT observations.   The field is not aligned  with the major axis of the pillar structure, and  differs by about 45 degrees from that in the HII region.  As discussed above, we hypothesize that SMM2 is embedded within the pillar and has inherited its magnetic field from the pre-shock molecular cloud, although we cannot rule out the scenario in which the core has been forced into collapse by  photon pressure or gas flows due to the formation of the pillar.

\section{Conclusions} \label{sec:conclusions}

We have presented the first polarized dust emission observations of the Horsehead Nebula, which were obtained using JCMT POL-2/SCUBA-2 at 850 $\mu$m. We also presented JCMT HARP C$^{18}$O $J=3\to2$ observations. We observed plane-of-sky magnetic field orientations in the SMM1 PDR and the SMM2 starless core in the Horsehead Nebula, in both cases seeing well-ordered magnetic fields aligned approximately perpendicular to the major axes of the sources.

We used the angular dispersion function variation on the Davis-Chandrasekhar-Fermi method to estimate plane-of-sky magnetic field strengths of 56$\pm$9 and 129$\pm$21 $\mu$G in SMM1 and SMM2, respectively. These correspond to sub-critical mass-to-flux ratios of 0.5$\pm$0.1 in SMM1 and 0.4$\pm$0.1 in SMM2 when we assume $B \sim B_{pos}$, indicating that magnetic fields can support the structures against gravitational collapse. We estimate Alfv\'en Mach numbers of 0.27$\pm$0.05 and 0.40$\pm$0.06 in SMM1 and SMM2, respectively, indicating that turbulence in these structures is sub-Alfv\'enic. We estimated the number of turbulent eddies along the line of sight, and accounted for this in these estimates of mass-to-flux ratio and Alfv\'en Mach number. In both regions, the magnetic field appears dynamically important compared to self-gravity and internal turbulent pressure.  However, we expect the energetics of the SMM1 PDR to be dominated by gas and photon pressure from the HII region with which it is interacting.

SMM2 has a magnetic field strength and geometry which is quite typical for a starless core, and which appears to have been inherited from the pre-shock molecular cloud in which the core is embedded. We estimated the energy balance in SMM2, finding that its magnetic energy is  approximately three times larger than its internal turbulent and gravitational potential energies, which are comparable to one another. The large magnetic energy in SMM2 could be caused by overestimation of the magnetic field strength by the DCF method. If the magnetic field energy is indeed dominant in SMM2, this suggests that the core is confined by external pressure.

The magnetic field which we observe in the SMM1 PDR appears to be significantly compressed and reordered by its interaction with the HII region, but to have developed from the same magnetic field geometry in the pre-shock molecular cloud which is seen, apparently undisturbed, in SMM2. The projected magnetic field geometry which we observe in SMM1 is consistent with a three-dimensional magnetic field aligned plane-parallel to the interface with the HII region if the magnetic field is either folded back on itself along the line of sight, or if we are observing the magnetic field in the plane of a PDR which is highly inclined along the line of sight.

\begin{acknowledgements}

J.H. is supported by the UST Overseas Training Program 2022, funded by the University of Science and Technology, Korea (No. 2022-017).  K.P. is a Royal Society University Research Fellow, supported by grant number URF\textbackslash R1\textbackslash 211322. The authors thank Janik Karoly for sharing his POL-2 pixel size analysis results ahead of publication. The JCMT data presented in this paper was awarded under the Maunakea Scholars program. Maunakea Scholars is an innovative program designed to bring Hawaii’s aspiring young astronomers into the observatory community, competitively allocating observing time on world-class telescopes to local students. In particular the authors wish to thank Doug Simons and Mary Beth Laychak as leads of the Maunakea Scholars program, and Paul Ho for his award of JCMT Director's Discretionary Time (DDT) to this project. More information can be found at: \url{https://maunakeascholars.com}. The JCMT is operated by the East Asian Observatory on behalf of the National Astronomical Observatory of Japan; the Academia Sinica Institute of Astronomy and Astrophysics; the Korea Astronomy and Space Science Institute; the Operation, Maintenance and Upgrading Fund for Astronomical Telescopes and Facility Instruments, budgeted from the Ministry of Finance of China. Additional funding support is provided by the Science and Technology Facilities Council of the United Kingdom and participating universities and organizations in the United Kingdom, Canada and Ireland.  Additional funds for the construction of SCUBA-2 were provided by the Canada Foundation for Innovation. The authors wish to recognize and acknowledge the very significant cultural role and reverence that the summit of Maunakea has always had within the indigenous Hawaiian community.  We are most fortunate to have the opportunity to conduct observations from this mountain.

\end{acknowledgements}






\begin{thebibliography}{}

\bibitem[Abergel et al.(2003)]{Abergel2003} Abergel, A., Teyssier, D., Bernard, J.~P., et al.\ 2003, \aap, 410, 577. doi:10.1051/0004-6361:20030878

\bibitem[Andr{\'e} et al.(2010)]{Andre2010} Andr{\'e}, P., Men'shchikov, A., Bontemps, S., et al.\ 2010, \aap, 518, L102

\bibitem[Arthur et al.(2011)]{Arthur2011} Arthur, S.~J., Henney, W.~J., Mellema, G., et al.\ 2011, \mnras, 414, 1747. doi:10.1111/j.1365-2966.2011.18507.x

\bibitem[Barnard(1919)]{Barnard1919} Barnard, E.~E.\ 1919, \apj, 49, 1. doi:10.1086/142439

\bibitem[Beckwith et al.(1990)]{Beckwith1990} Beckwith, S.~V.~W., Sargent, A.~I., Chini, R.~S., et al.\ 1990, \aj, 99, 924

\bibitem[Chandrasekhar \& Fermi(1953)]{ChanFer1953} Chandrasekhar, S., \& Fermi, E.\ 1953, \apj, 118, 113

\bibitem[Chapin et al.(2013)]{Chapin2013} Chapin, E.~L., Berry, D.~S., Gibb, A.~G., et al.\ 2013, \mnras, 430, 2545. doi:10.1093/mnras/stt052

\bibitem[Cho \& Yoo(2016)]{Cho2016} Cho, J. \& Yoo, H.\ 2016, \apj, 821, 21. doi:10.3847/0004-637X/821/1/21

\bibitem[Coud\'e et al.(2019)]{Coude2019} {Coud{\'e}}, S., {Bastien}, P. {Houde}, M., et al.\ 2019, \apj, 877, 88. doi:10.3847/1538-4357/ab1b23

\bibitem[Crutcher et al.(1999)]{Crutcher1999} Crutcher, R.~M., Troland, T.~H., Lazareff, B., et al.\ 1999, \apjl, 514, L121

\bibitem[Crutcher(2004a)]{Crutcher2004a} Crutcher, R.~M.\ 2004, \apss, 292, 225. doi:10.1023/B:ASTR.0000045021.42255.95

\bibitem[Crutcher et al.(2004b)]{Crutcher2004b} Crutcher, R.~M., Nutter, D.~J., Ward-Thompson, D., et al.\ 2004, \apj, 600, 279. doi:10.1086/379705

\bibitem[Currie et al.(2008)]{Currie2008} Currie, M.~J., Draper, P.~W., Berry, D.~S., et al.\ 2008, Astronomical Data Analysis Software and Systems XVII, 394, 650

\bibitem[Currie et al.(2014)]{Currie2014} Currie, M.~J., Berry, D.~S., Jenness, T., et al.\ 2014, Astronomical Data Analysis Software and Systems XXIII, 485, 391

\bibitem[Davis(1951)]{Davis1951} Davis, L.\ 1951, Physical Review, 81, 890


\bibitem[Dempsey et al.(2013)]{Dempsey2013} {Dempsey}, J.~T., {Friberg}, P., {Jenness}, T., et al.\ 2013, \mnras, 430, 2534. doi:10.1093/mnras/stt090

\bibitem[Fiedler \& Mouschovias(1993)]{Fiedler1993} {Fiedler}, R. A. \& {Mouschovias}, T. C.\ 1993, \apj, 415 680. doi:10.1086/173193

\bibitem[Friberg et al.(2016)]{Friberg2016} {Friberg}, P., {Bastien}, P., {Berry}, D., et al.\ 2016, Proc. SPIE, 9914, 991403. doi:10.1117/12.2231943


\bibitem[Guerra et al.(2021)]{Guerra2021} {Guerra}, J. A., {Chuss}, D. T., {Dowell}, C. D. et al\ 2021, \apj, 908, 98. doi:10.3847/1538-4357/abd6f0

\bibitem[Habart et al.(2005)]{Habart2005} Habart, E., Abergel, A., Walmsley, C.~M., et al.\ 2005, \aap, 437, 177. doi:10.1051/0004-6361:20041546

\bibitem[Henney et al.(2009)]{Henney2009} {Henney}, W. J., {Arthur}, S. J., {de Colle}, F. et al.\ 2009, \mnras, 398, 157. doi:10.1111/j.1365-2966.2009.15153.x

\bibitem[Hildebrand et al.(2009)]{Hildebrand2009} Hildebrand, R.~H., Kirby, L., Dotson, J.~L., et al.\ 2009, \apj, 696, 567. doi:10.1088/0004-637X/696/1/567

\bibitem[Hildebrand(1983)]{Hildebrand1983} Hildebrand, R.~H.\ 1983, \qjras, 24, 267

\bibitem[Hiltner(1949)]{Hiltner1949} Hiltner, W.~A.\ 1949, Science, 109, 165. doi:10.1126/science.109.2825.165

\bibitem[Holland et al.(2013)]{Holland2013} Holland, W.~S., Bintley, D., Chapin, E.~L., et al.\ 2013, \mnras, 430, 2513. doi:10.1093/mnras/sts612

\bibitem[Jenness et al.(2013)]{Jenness2013} Jenness, T., Chapin, E.~L., Berry, D.~S., et al.\ 2013, Astrophysics Source Code Library, ascl:1310.007 

\bibitem[Karoly et al.(2020)]{Karoly2020} Karoly, J., Soam, A., Andersson, B.-G., et al.\ 2020, \apj, 900, 181. doi:10.3847/1538-4357/abad37

\bibitem[Kauffmann et al.(2008)]{Kauffmann2008} Kauffmann, J., Bertoldi, F., Bourke, T.~L., et al.\ 2008, \aap, 487, 993

\bibitem[Kirk et al.(2006)]{Kirk2006} {Kirk}, J.~M., {Ward-Thompson}, D. \& {Crutcher}, R.~M.\ 2006, \mnras, 369, 1445. doi:10.1111/j.1365-2966.2006.10392.x

\bibitem[K{\"o}nyves et al.(2021)]{Konyves2021} K{\"o}nyves, V., Ward-Thompson, D., Pattle, K., et al.\ 2021, \apj, 913, 57. doi:10.3847/1538-4357/abf3ca

\bibitem[Lazarian \& Hoang(2007)]{Lazarian2007} Lazarian, A. \& Hoang, T.\ 2007, \mnras, 378, 910

\bibitem[Liu et al.(2019)]{Liu2019} Liu, J., Qiu, K., Berry, D., et al.\ 2019, \apj, 877, 43. doi:10.3847/1538-4357/ab0958

\bibitem[Liu et al.(2021)]{Liu2021} {Liu}, J., {Zhang}, Q., {Commer{\c{c}}on}, B., et al.\ 2021, \apj, 919, 79. doi:10.3847/1538-4357/ac0cec

\bibitem[Mairs et al.(2021)]{Mairs2021} Mairs, S., Dempsey, J.~T., Bell, G.~S., et al.\ 2021, \aj, 162, 191. doi:10.3847/1538-3881/ac18bf

\bibitem[Motte \& Andr{\'e}(2001)]{MotteAndr2001} Motte, F., \& Andr{\'e}, P.\ 2001, \aap, 365, 440 

\bibitem[Mouschovias \& Spitzer(1976)]{Mouschovias1976} Mouschovias, T.~C., \& Spitzer, L.\ 1976, \apj, 210, 326

\bibitem[Nakano \& Nakamura(1978)]{Nakano1978} Nakano, T. \& Nakamura, T.\ 1978, \pasj, 30, 671

\bibitem[Pattle et al.(2022)]{Pattle2022}
{Pattle}, K., {Fissel}, L., {Tahani}, M. et al.\ 2022, Protostars \& Planets 7 Proceedings, arXiv:2203.11179

\bibitem[Pattle et al.(2021)]{Pattle2021} Pattle, K., Lai, S.-P., Di Francesco, J., et al.\ 2021, \apj, 907, 88. doi:10.3847/1538-4357/abcc6c

\bibitem[Pattle \& Fissel(2019)]{Pattle2019} Pattle, K. \& Fissel, L.\ 2019, Frontiers in Astronomy and Space Sciences, 6, 15. doi:10.3389/fspas.2019.00015

\bibitem[Pattle et al.(2018)]{Pattle2018} Pattle, K., Ward-Thompson, D., Hasegawa, T., et al.\ 2018, \apjl, 860, L6. doi:10.3847/2041-8213/aac771

\bibitem[Pattle et al.(2017)]{Pattle2017} Pattle, K., Ward-Thompson, D., Berry, D., et al.\ 2017, \apj, 846, 122. doi:10.3847/1538-4357/aa80e5

\bibitem[Pattle et al.(2015)]{Pattle2015} {Pattle}, K., {Ward-Thompson}, D., {Kirk}, J.~M., et al.\ 2015, \mnras, 450, 1094. doi:10.1093/mnras/stv376

\bibitem[Planck Collaboration et al.(2015)]{Planck2015} Planck Collaboration, Ade, P.~A.~R., Aghanim, N., et al.\ 2015, \aap, 576, A104. doi:10.1051/0004-6361/201424082

\bibitem[Pound et al.(2003)]{Pound2003} Pound, M.~W., Reipurth, B., \& Bally, J.\ 2003, \aj, 125, 2108. doi:10.1086/368138

\bibitem[Rigby et al.(2016)]{Rigby2016}
{Rigby}, A.~J., {Moore}, T.~J.~T., {Plume}, R. et al.\ 2016, \mnras, 456, 2885. doi:10.1093/mnras/stv2808

\bibitem[Schaefer et al.(2016)]{Schaefer2016}
{Schaefer}, G.~H., {Hummel}, C.~A., {Gies}, D.~R., et al.,\ 2016, \aj, 152, 213. doi:10.3847/0004-6256/152/6/213

\bibitem[Walch et al.(2012)]{Walch2012} Walch, S.~K., Whitworth, A.~P., Bisbas, T., et al.\ 2012, \mnras, 427, 625. doi:10.1111/j.1365-2966.2012.21767.x

\bibitem[Ward-Thompson et al.(2000)]{WardThompson2000}
{Ward-Thompson}, D., {Kirk}, J.~M., {Crutcher}, R.~M. et al.\ 2000, \apjl, 537, L135. doi:10.1086/312764

\bibitem[Ward-Thompson et al.(2006)]{WardThompson2006} Ward-Thompson, D., Nutter, D., Bontemps, S., et al.\ 2006, \mnras, 369, 1201. doi:10.1111/j.1365-2966.2006.10356.x

\bibitem[Ward-Thompson et al.(2017)]{WardThompson2017} {Ward-Thompson}, D., {Pattle}, K., {Bastien}, P. et al.\ 2017, \apj, 842, 66. doi:10.3847/1538-4357/aa70a0

\bibitem[Zaritsky et al.(1987)]{Zaritsky1987} Zaritsky, D., Shaya, E.~J., Scoville, N.~Z., et al.\ 1987, \aj, 93, 1514. doi:10.1086/114432

\end{thebibliography}
\end{document}